 \documentclass[final,5p,times,twocolumn]{elsarticle}

\usepackage{amssymb}
\usepackage{lipsum}
\usepackage{amsmath}

\newcommand{\ion}[2]{\text{#1\,\uppercase\expandafter{\romannumeral #2}}}

\usepackage{hyperref}

\journal{JQSRT}
\begin{document}

\begin{frontmatter}



\title{The Valence and Inner Shell Ionization Properties of \ion{Sr}{1}~-\ion{}{3} by Non-Thermal Electrons in Kilonovae}


\author[first]{S. J. Bromley*}
\affiliation[first]{organization={Department of Physics, Auburn University},
addressline={Leach Science Center},
city={Auburn},
postcode={36849},
state={AL},
country={USA}}
\cortext[cor1]{Corresponding author}
\ead{sjb0068@auburn.edu}

\author[second]{L. Shingles}
\affiliation[second]{organization={Astrophysics Research Centre, School of Mathematics \& Physics, Queen's University Belfast},
addressline={},
city={Belfast},
postcode={BT7 1NN},
state={},
country={Northern Ireland}}

\author[third]{E. Garbe}
\affiliation[third]{organization={Department of Physics and Astronomy and the Center for Simulational Physics, University of Georgia},
addressline={},
city={Athens},
postcode={30602},
state={GA},
country={USA}}

\author[second]{N. McElroy}

\author[second]{C.P. Ballance}

\author[first]{M. Fogle}

\author[third]{P. C. Stancil}

\author[first]{S. D. Loch}
\begin{abstract}
The presence of strontium ($Z = 38$) was identified from spectra of the first multi-messenger kilonova transient, AT2017gfo, with the evolution of its spectral signature indicating a rapidly changing ionization balance. Interpretation of spectra at later times requires a more detailed understanding of recombination and ionization processes, including ionization by `non-thermal' electrons produced by beta decay. Published measurements of the ionization cross sections of \ion{Sr}{1} and \ion{Sr}{2} show complex structures likely arising from excitation autoionization. Using the Flexible Atomic Code we report calculations of the valence and inner shell ionization cross sections for \ion{Sr}{1}, \ion{Sr}{2}, and \ion{Sr}{3}, and recover similar structures in the ionization cross sections. Though no measurements exist for \ion{Sr}{3}, the calculations suggest negligible contributions from excitation autoionization. These new data are incorporated into Spencer-Fano models of non-thermal electron degradation and coupled to an ionization balance solver. The effects of excitation autoionization, treated as excitation with the additional production of an Auger electron, on the degradation spectra of non-thermal are explored. Calculations at both level- and configuration-average resolution return ion fractions that agree within 5-20\%. The calculations probe the impact of including final-charge-state-resolved branching ratios for inner shell ionization which result in multi-ionization events from successive Auger cascades. Omission of these branching ratios is shown to have large impacts on the ionization balance. The calculated cross sections will be provided on Zenodo.
\end{abstract}

\end{frontmatter}



\section{Introduction}\label{sec:intro}
Strontium ($Z = 38$) is formed in part by $r$-process nucleosynthesis and is one of the earliest identified elements in spectra of AT2017gfo~\citep{Watson2019}, the transient which followed from the first multi-messenger detection of a neutron star, GW170817~\citep{Abbott2017}. An analysis of the time-series spectra of AT2017gfo by Sneppen \textit{et al.}~\cite{Sneppen2024} identified an evolving absorption feature around 1~$\mu$m, from which they inferred a rapidly-changing ionization balance of Sr ions. While such early times clearly motivate an opacity-based analysis built on Local Thermodynamic Equilibrium (LTE) and Saha equilibrium (see for example the works of Refs.~\cite{Kasen2013,Hotokezaka2018,Fontes2022,Gillanders2024,Vieira_2023}), analysis beyond the LTE-like photospheric phase requires an understanding of the ionization and recombination properties of ions~\cite{Pognan2021,Pognan2023,Jerkstrand2025}.

Over time both the radiation field and particle density within kilonovae ejecta are increasingly diluted, and the originally-high densities motivating Saha/LTE treatments in the photospheric phase are no longer valid. Instead, non-equilibrium models must be applied, and require the inclusion of all processes that affect the level populations and ionization balance: fluorescence, recombination, and electron impact excitation and ionization~\cite{Pognan2021,PognanNLTE,Hotokezaka2022,Pognan2023,Jerkstrand2025}. These electron impact collisions are mediated by two populations of electrons. First, the bulk of the electrons can be considered `thermal', and are described by a Maxwellian distribution at temperatures of order few thousand~Kelvin. Recombination proceeds through dielectronic and radiative recombination with these thermal electrons. The requisite atomic data describing this process are temperature-dependent, Maxwellian-averaged rate coefficients. Second, `non-thermal' electrons, the products of beta decays, are initially formed at energies spanning the range keV to MeV, and continually lose energy through Coulomb collisions with thermal electrons, bound-bound, and bound-free collision with the ejecta~\citep{Barnes_2016}. Beyond the optically-thick `photospheric' LTE phase of a kilonova, electron impact ionization by non-thermal electrons contributes significantly to and possibly dominates the ionization rates~\cite{brethauer2025nonthermalionizationkilonovaejecta}. The distribution of non-thermal electron energies however depends on the energy released by beta decay(s) within the ejecta and the associated electron impact cross sections for the ions present. Computationally, the degradation of these electrons as they interact with the plasma - and the energy deposition through ionization - is encapsulated within the Spencer-Fano formalism~\citep{Spencer1954,Kozma1992}.

Due to limited applications beyond kilonovae, many of the $r$-process elements of interest to kilonovae have received little to no experimental or theoretical attention in regards to their ionization and recombination properties. In the absence of explicit ionization data, it is beneficial for the purposes of ionization balance modeling to approximate their ionization cross sections with the expression of Lotz~\cite{Lotz1967}. This expression, and its empirical scaling factor, was originally derived from a limited number of measurements of neutral or near-neutral ions. As a consequence there exists large uncertainties on computed ion fractions in kilonovae ejecta. However, measurements of the single ionization cross sections of \ion{Sr}{1}~\cite{Vainshtein1972,Okudaira1970} and \ion{Sr}{2}~\cite{Peart1975,Peart1989} are available and show large contributions from excitation autoionization, a process which is absent within the Lotz formalism. Additionally, non-thermal electrons produced from beta decays in kilonovae ejecta carry sufficiently high energies that ionization can also proceed through the production of inner shell holes. It is presently unknown how the inclusion or exclusion of these effects impact ionization balance models of Sr ions at kilonova-like conditions.

To understand these effects, Flexible Atomic Code calculations of the ionization cross sections of \ion{Sr}{1}~-\ion{}{3} were carried out, and these data were incorporated into Spencer-Fano models of non-thermal electron degradation with self-consistent ionization balance modeling. Particular attention is paid toward the contributions of excitation autoionization and inner shell ionization. The calculations are performed at both level- and configuration-average resolution. To evaluate the level of improvement in both electron degradation and ionization balance modeling, these data are also compared to empirical estimates of excitation and ionization, and followed through to complete Spencer-Fano/ionization balance modeling using all generated cross section datasets.

This manuscript proceeds as follows. The application of the Flexible Atomic Code and the full suite of cross section calculations is described in Sections~\ref{sec:methods} and \ref{sec:results}. The incorporation of the cross sections into a model of non-thermal electron degradation is described in Section~\ref{sec:nonthermal}, and used to probe the differences in self-consistent ionization balance models generated from different cross section datasets. The context of these calculations, and their implications on future studies of ionization balances in kilonovae, is discussed in Sec.~\ref{sec:discussion}. The findings are summarized in Section~\ref{sec:conclusions}.

\section{Cross Section Calculations with the Flexible Atomic Code}\label{sec:methods}
To compute the ionization properties of Sr ions we employ the Flexible Atomic Code (FAC;~\cite{FAC}) with an approach similar to previously reported calculations for Te ions~\citep{Bromley_2026}. FAC is a fully-relativistic atomic structure package equipped to compute energy levels, bound-bound, and bound-free  collisional properties of atomic ions. Both the Binary Encounter Dipole and Distorted Wave approximations are used for calculations of direct ionization cross sections. The Distorted Wave formalism is also used for calculations of excitation cross sections in order to evaluate the role of excitation autoionization.

Compared to a `typical' Maxwellian plasma where inner shell ionization is weak, non-thermal electrons in kilonovae carry energies that are typically much larger than the single ionization potential, and ionization proceeds by removal of both valence and inner shell electrons. Compared to deeply bound electrons in inner shells, valence electrons are more strongly affected by configuration interaction. We therefore separate out the FAC calculations into two parts: first, for the purposes of computing single ionization cross sections of the valence electron(s), a list of electronic configurations was generated from single and double promotions out of the known ground state of \ion{Sr}{1}, 4p$^6$5s$^2$, where closed shells are omitted in this labeling scheme. This procedure yields a list of configurations that are likely to mix with levels of interest through configuration interaction, and encompass the configurations which could potentially contribute through excitation autoionization. These configurations consist of all unique combinations of the following forms: 4p$^6$ (5s,5p,4d) $n\ell$ ($n \leq20$, $\ell \leq 4$), 4p$^5$ 5s (5s,5p,4d) $n\ell$ ($n \leq20$, $\ell \leq 4$), and 4p$^5$ (4d$^2$,5s$^2$,5p$^2$) $n\ell$ ($n \leq20$, $\ell \leq 4$). The process was repeated for \ion{Sr}{2} and \ion{Sr}{3} which yielded similar forms of the electronic configurations.

Within the FAC, all properties are computed in the presence of a central field potential common to all levels within both the initial (e.g. \ion{Sr}{1}) and final (e.g. \ion{Sr}{2}) ions. The optimal central field potential is generally found by minimizing the energy of a fictitious `mean' electronic configuration, here chosen as the ground state of the initial ion. Test calculations optimized on the ionized system within each pair of ions, such as \ion{Sr}{1} and \ion{Sr}{2}, yielded ionization cross sections a factor of $\sim$2 smaller than the available measurements and were not explored further. Test calculations were also performed to probe whether limiting the configuration interaction between levels affected the cross sections. Limiting the configuration interaction to levels within each relativistic or non-relativistic configuration lead to cross sections that agreed poorly with measurement. We therefore adopt a central potential optimized on the ground state of the initial ion, and allowed all levels within each parity to mix via configuration interaction.

The calculations presented herein were performed in both full level-resolved detail, i.e. levels are resolved in terms of their total angular momenta $J$, and in the configuration average (CA) approximation. In a typical CA approach, the energy levels of a given electronic configuration are averaged into a representative level with an appropriate energy and statistical weight. Within the relativistic FAC, the CA approach averages levels over relativistic sub-shells. While the resulting CA energy level structure is not at spectroscopic accuracy, the number of energy levels in the calculations is significantly reduced. This decreases the total number of rates and cross sections required for the calculations of e.g. excitation autoionization, and lowers the associated computational costs. 

The full suite of calculations proceeded as follows. For each ion stage, the energy levels of the initial and ionized systems were computed in the presence of a common central field potential optimized on the ground state of the lower charge state. Ground and metastable levels (in the calculations involving \ion{Sr}{2} and \ion{Sr}{4}), as well as inner shell excitation thresholds, were shifted to known values where possible~\citep{NIST_ASD}. Direct ionization cross sections were computed in both the Binary Encounter Dipole and Distorted Wave approximations. For both \ion{Sr}{1} and \ion{Sr}{2}, the Binary Encounter Dipole approximation was found to produce a cross section that agreed well with experiment. Excitation cross sections were computed in the Distorted Wave approximation and converted to excitation autoionization (EAI) cross sections using an appropriate branching ratio computed from a cascade calculation. Nearly all levels located above the relevant ionization potential have a  branching ratio close to unity; regardless, we utilize the explicitly-computed branching ratio in the calculations of the excitation-autoionization cross sections. Last, we removed all EAI contributions from known bound levels by comparisons of the configuration labels and the energies computed with FAC against the established energy levels in the NIST Atomic Spectra Database~\citep{NIST_ASD}.

\begin{figure*}[t]
        \centering
   \includegraphics[scale=0.6]{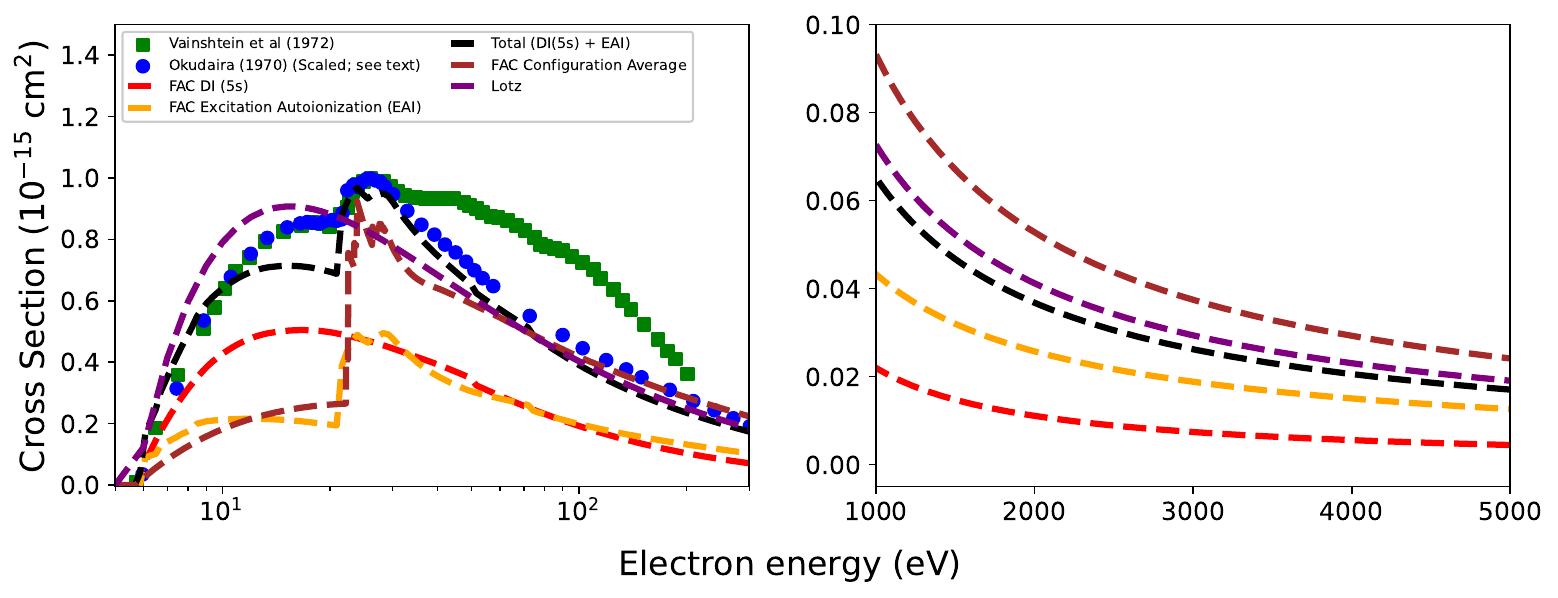}
\caption{\textit{(Left)} Comparison of the computed cross sections and the measurements of Vainshtein \textit{et al.} \cite{Vainshtein1972} and Okudaira~\cite{Okudaira1970} on a logarithmic energy scale. The presented cross sections of Ref.~\cite{Okudaira1970} are relative cross sections which were scaled to the maximum of the cross section reported in Ref.~\cite{Vainshtein1972}. Differences between the two experiments and the theoretical calculations are likely due to multi-ionization processes (see text). \textit{(Right)} The high energy tails (on a linear energy scale) highlight differences in the cross sections at the asymptotic limit, chosen here as 5 keV.}\label{fig:sr0_cs}
\end{figure*}

Second, for inner shell ionization, we consider ionization of all sub-shells with binding energies greater than the second ionization potential of the initial ion, with the exception of the $1s$ orbital, which we omit due to its negligible ionization cross section compared to the total ionization cross section. Direct ionization calculations of inner shells were computed using the perturbative Distorted Wave approximation. For $K$-shell ionization, relativistic Distorted Wave approaches have shown good agreement with experiment~\citep{Colgan2006}, particularly for neutral atoms~\citep{BOTE2009871}. For $L$ and $M$ shells, similar levels of accuracy are expected owing to the strongly bound nature of these shells and their relative insensitivity to many-body effects. In the case of \ion{Sr}{1}, removal of any electron in the $4p$ sub-shell or below results in at least a double ionization due to Auger cascades. In \ion{Sr}{2}, the same follows from removal of $4s$ electrons and below; in \ion{Sr}{3}, removal of electrons from the $3d$ shell and below is expected to produce multi-ionization events, based on comparisons of the binding energies and the known ionization potentials. 

\section{Cross Section Results}\label{sec:results}
\subsection{Single Ionization: \ion{Sr}{1}}
The computed single ionization cross sections for \ion{Sr}{1} are shown in Figure~\ref{fig:sr0_cs} where the calculations are compared to measurements by  Vainshtein \textit{et al.} \cite{Vainshtein1972} and Okudaira \cite{Okudaira1970}. The cross sections presented in Ref.~\cite{Okudaira1970} are relative cross sections, which are scaled here to the peak of the cross section reported in Ref.~\cite{Vainshtein1972}. Comparing the two measurements, Ref.~\cite{Okudaira1970} explicitly separated multi-ionization events from single ionization events, while the apparatus of Ref.~\cite{Vainshtein1972} appears to have collected all ions, resulting in a total (and not single) ionization cross section. The present cross section agrees well in the threshold region with the work of Ref.~\cite{Vainshtein1972}, but differs by $10-15\%$ between 11 eV and the $4p$ excitation threshold around 21 eV. However, the computed cross section agrees favorably with the shape of the presumed single ionization cross section reported by Ref.~\cite{Okudaira1970}, particularly toward the higher energies. Differences between the present work and the measurements of Vainshtein \textit{et al.}~\cite{Vainshtein1972} can be attributed to multi-ionization processes which are excluded from the present calculations.

The apparent contributions from excitation-autoionization in the measurements can be attributed to several distinct channels included in the calculations. The majority of the autoionization contributions below 21 eV are attributed to the doubly-excited configurations $4p^64d6p$ and $4p^65p6s$, the former of which have known energies just above the ionization potential of \ion{Sr}{1}~\citep{NIST_ASD}. Above 21 eV, the excitation autoionization cross section is dominated by inner-shell excitations to the configurations $4p^55s^24d$ and $4p^54d^25s$, both of which exhibit low purity due to strong mixing.

The FAC cross sections are also compared with cross sections estimated by the Lotz \cite{Lotz1967} approximation which take the form
\begin{equation}\label{eq:lotz}
    \sigma_{Lotz}(E) = \sum_i  n_i a_i \frac{\textrm{ln}(E/I_i)}{EI_i}
\end{equation}
where $i$ is a sum over the $n\ell$ shells, $n_i$ is the number of electrons in each shell, $a$ is a constant, and $I_i$ is the ionization potential of the $i$ shell. We use a nominal value of $a$ = 4$\times$10$^{-14}$~cm$^2$~eV$^2$, taken from fits to \ion{Rb}{1} cross sections reported in Ref.~\cite{Lotz1967}. For \ion{Sr}{1}, we set $n_i$ = 2 for $5s$ ionization and $n_i$ = 6 for $4p$ ionization. Appropriately adjusted values of the occupation numbers $n_i$ were utilized for later ion stages. The Lotz ionization cross section agrees well near the ionization threshold but fails to account for the enhanced cross section due to excitation-autoionization. 

\begin{figure*}
        \centering
   \includegraphics[scale=0.6]{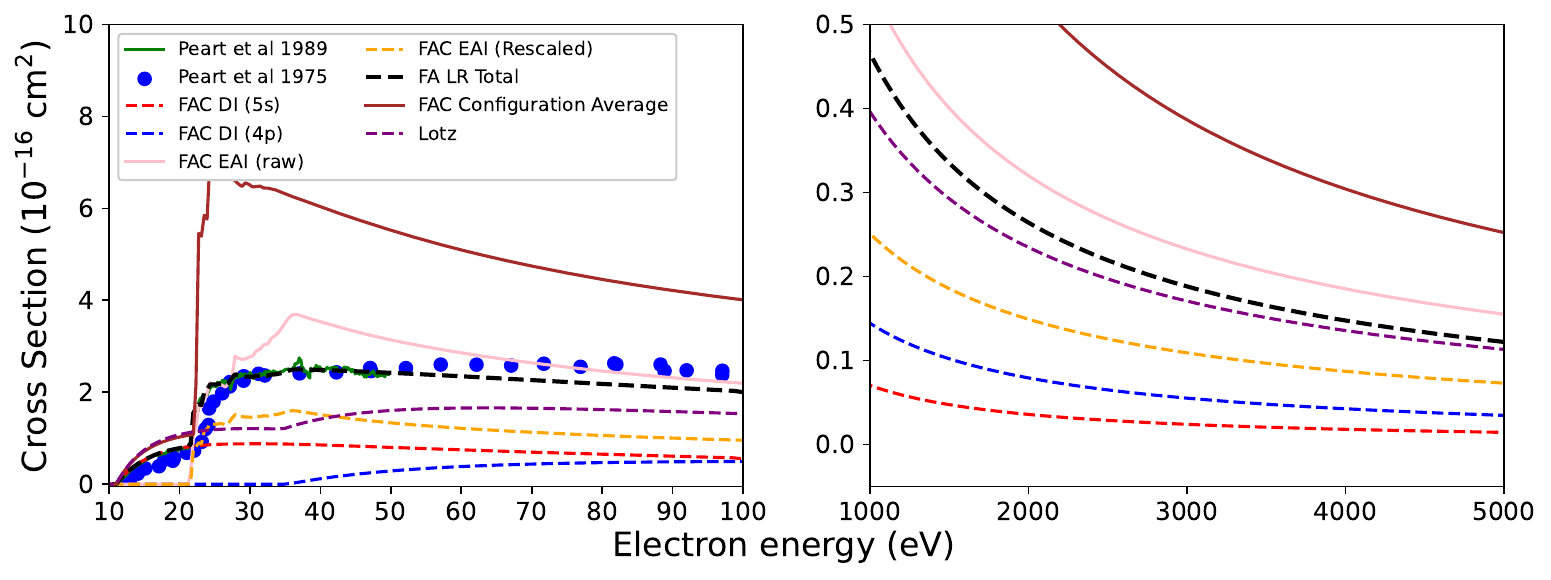}
\caption{\textit{(Left)} Comparison of the present FAC calculations and the measurements of Peart \textit{et al.}~\cite{Peart1975,Peart1989} in the threshold region. Measurements were digitized from published figures. The total cross section from the level-resolved calculations (black) is computed as the sum of both $5s$ and $4p$ direct ionization, and the sum of the manually-adjusted excitation-autoionization channels. Raw excitation autoionization data restricted to states with threshold energies above 21.3 eV is shown for comparison. The FAC configuration average single ionization cross section is computed as the sum of direct ionization and excitation autoionization, the latter without any rescaling and only a requirement that EAI thresholds are greater than the known $4p$ excitation threshold. \textit{(Right)} The high-energy behavior of the various cross sections.}\label{fig:sr1_cs}
\end{figure*}

Figure~\ref{fig:sr0_cs} also shows single ionization cross sections computed in the configuration average (CA) approximation. The CA cross section underestimates the direct ionization contribution, but recovers a similar EAI structure as the level-resolved results. At higher energies above 100 eV, the magnitude of the CA cross section is $\sim$30\% larger than the level-resolved result. Though the magnitudes of the CA cross sections are low near threshold, they do indicate that excitation autoionization is, as expected from both the level-resolved calculations and the available measurements, a sizable contribution to the ionization cross section. 

Lastly, Figure~\ref{fig:sr0_cs} (right) also shows the asymptotic behavior of the computed and empirical cross sections. As will be discussed later, the non-thermal ionization rate in kilonovae is proportional to both the non-thermal electron flux and the energy-dependent cross sections. Direct ionization provides only $\sim$half of the total ionization cross section. The remaining fraction is provided by excitation autoionization, which scales identically to the ionization cross section as $\textrm{ln}(E)/E$ at large energies. Though the threshold region contributes negligibly to the ionization rate by non-thermal electrons in kilonovae, matching experimentally-determined excitation autoionization features near and just above the threshold places the magnitude at the more-important higher energies on the right scale.

\subsection{Single Ionization: \ion{Sr}{2}}
The computed single ionization cross sections of \ion{Sr}{2} are shown in Figure~\ref{fig:sr1_cs} and compared to the measurements of Peart \textit{et al.}~\cite{Peart1975,Peart1989}. No structures in the cross section suggest the presence of metastable levels. Compared to \ion{Sr}{1}, \ion{Sr}{2} exhibits a more complex ionization cross section. Below $\sim$21.3 eV, the cross section can be attributed to direct $5s$ ionization, where the direct Binary Encounter Dipole ionization cross section is in excellent agreement with measurement. Above 21.3 eV, the direct ionization cross section is dwarfed by excitation autoionization. Within the raw EAI data several levels are erroneously located above the single ionization potential and the resulting total cross section (not shown) disagrees with measurement near threshold. These levels were removed from the EAI cross section in order for the total cross section to match the direct contribution beneath the $4p$ excitation threshold. Above this energy ($\sim$21.3 eV), the raw EAI data (pink curve) is in excess of the measurements of Peart \textit{et al.}

Attempts were made to vary the level of configuration interaction in the calculations, as well as the choice of configuration(s) used to optimize the central field potential, in order to study the sensitivity of the EAI channels to these user-specific optimization choices. This procedure was unable to provide a suitable match between the computed total ionization cross section and the measurements. For the purposes of Spencer-Fano and ionization balance modeling, the exact contributions from indirect ionization channels must be well-described. The direct ionization contribution (beneath the $4p$ excitation threshold) finds excellent with measurement, and we leave the magnitude of this cross section as-is. Above the $4p$ excitation threshold many of the structures present within the high-resolution cross section measurements of Peart \textit{et al.}~\cite{Peart1989} can be directly attributed to individual electronic configurations in the calculations. We therefore re-scaled the weight of each EAI channel to the total EAI cross section in order to match the observed structures within the measured ionization cross section. The weight of each configuration was varied by factors of $\pm$4, depending on the region of interest. The resulting total cross section is shown in Fig.~\ref{fig:sr1_cs} (black lines), where this `fit' procedure leads to a significant decrease in the EAI contributions (orange lines). The manual `fit', being the sum of both direct and indirect ionization cross sections, reproduces the experimental cross section of Ref.~\cite{Peart1989} within an average of $5.7\%$ above the $4p$ threshold. 

Considering also the two metastable levels of \ion{Sr}{2} belonging to the configuration $4p^64d$, the computed excitation-autoionization cross sections before any adjustments are nearly identical in shape (barring a $\sim$2 eV shift in threshold energies) to those of the ground state. Total cross sections for the two metastable levels were prepared in the same manner as the ground state using the same weights. The resulting ionization cross sections for the two metastable levels are $
\sim30\%$ smaller than their original values, though we expect little population in these levels even in kilonovae-like conditions. 

\begin{figure}[t]
        \centering
   \includegraphics[scale=0.6]{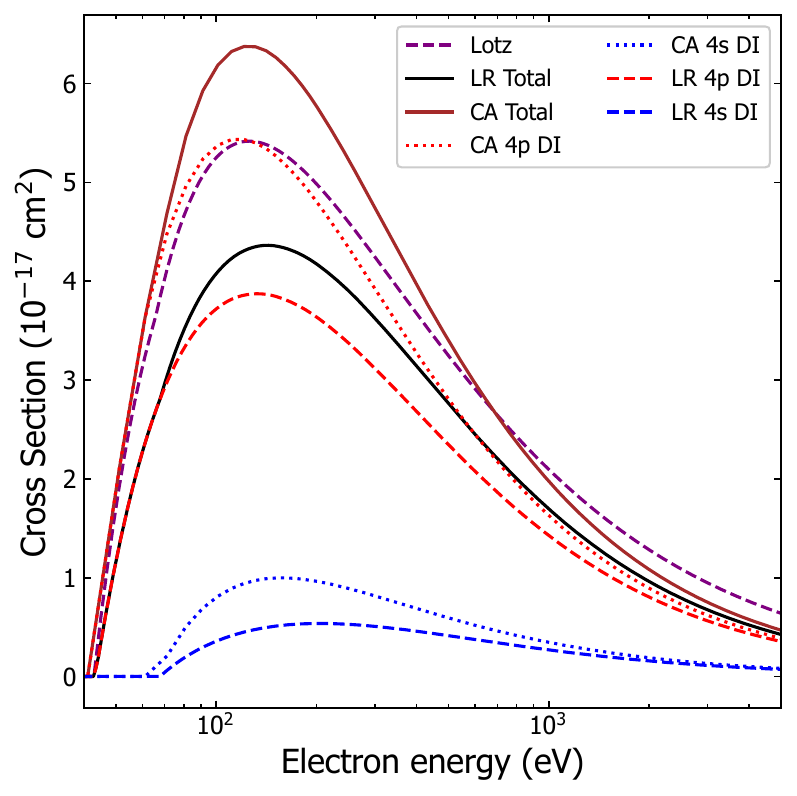}
\caption{Computed single ionization cross sections of \ion{Sr}{3}. All FAC-based cross sections (level-resolved, LR, and configuration average, CA) were computed with the Binary Encounter Dipole approximation. Lotz cross sections are computed as the sum of $4s$ and $4p$ ionization.}\label{fig:sr2_cs}
\end{figure}

Considering the same ionization processes at configuration average resolution, the CA approach underestimates the magnitude of the direct ionization cross section reported in Ref.~\cite{Peart1989}, but matches instead the direct ionization cross section from the earlier measurements of Ref.~\cite{Peart1975}. At CA resolution, the excitation autoionization cross sections are nearly a factor of three larger than the level-resolved cross sections. Using the same configuration-dependent weights from the level-resolved calculations, the total CA cross section can be made to agree with the measurements of Peart \textit{et al.} within $\sim$20\% for energies above the inner shell excitation threshold. However, we omit any adjustments of the configuration weights in the CA dataset in order to probe the impact of this overestimated EAI contribution on Spencer-Fano and ionization balance modeling in Section~\ref{sec:nonthermal}.

Inspection of Figure~\ref{fig:sr1_cs} shows that the Lotz cross section overestimates the direct ionization of both the $5s$ and $4p$ electrons. These overestimates offset in part the lack of excitation autoionization in the Lotz cross sections, resulting in an  asymptotic value that remains comparable to the level-resolved FAC result at 5 keV. 

\begin{figure*}[t]
        \centering
   \includegraphics[scale=0.57]{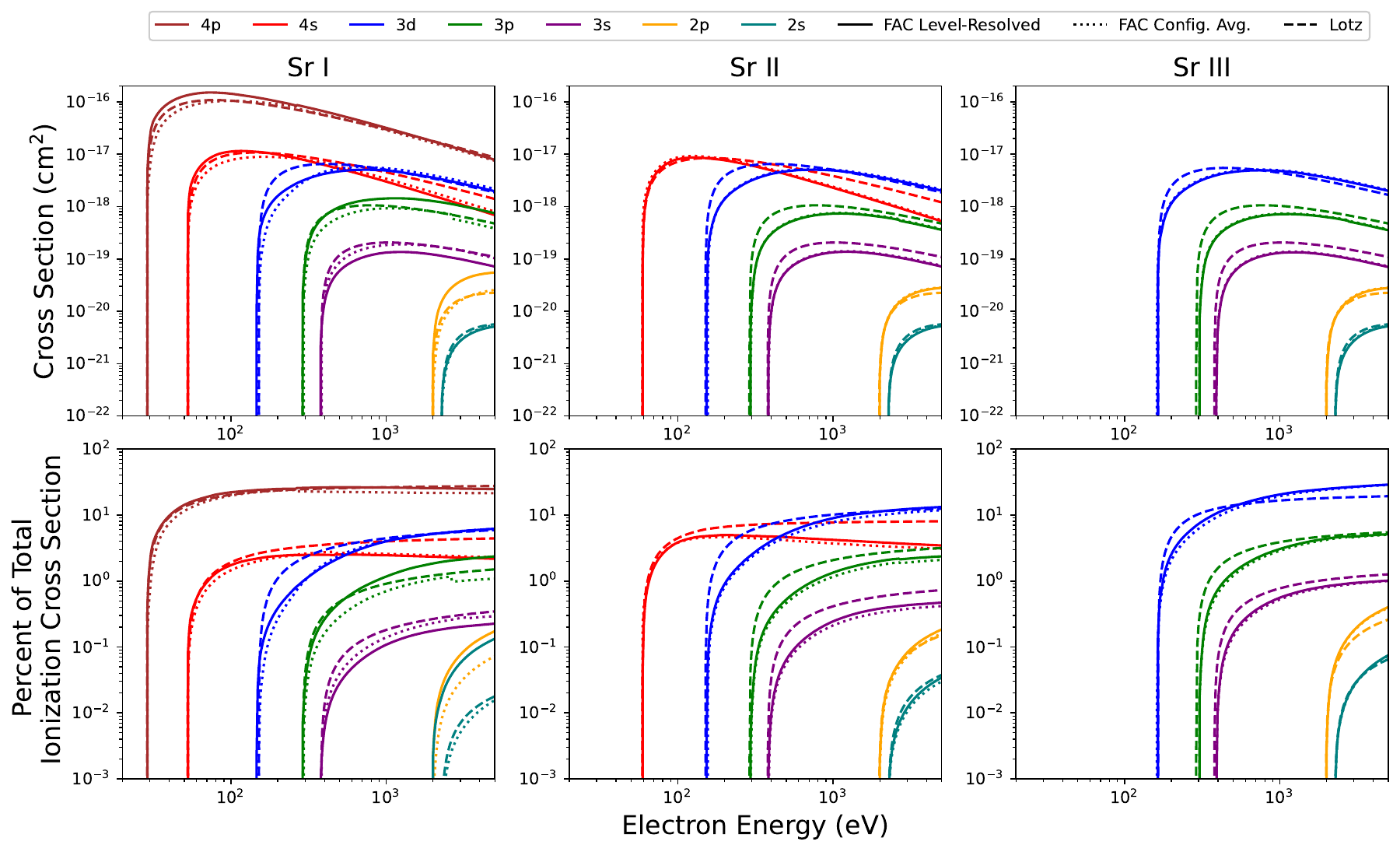}
\caption{(Top panel) Inner-shell ionization cross sections for \ion{Sr}{1} (left), \ion{Sr}{2} (center), and \ion{Sr}{3} (right). Calculations were performed in the Distorted Wave approximation in the FAC in full level-resolved detail (solid lines) and the configuration average approximation (dotted lines). Empirical Lotz cross sections were computed from Eq.~\ref{eq:lotz} with the empirical constant $a_i = 4\times10^{-14}$~cm$^2$~eV$^2$ and are shown as dashed lines. Colors indicate the hole formed in the ionization process. (Bottom Panel) Percent contribution of each inner shell channel to the total ionization cross section.}\label{fig:innershell_cs}
\end{figure*}

\subsection{Single Ionization: \ion{Sr}{3}}
Ionization of \ion{Sr}{3} proceeds through direct ionization of either the $4s$ or $4p$ electron in the ground configuration $4s^24p^6$. Ionization of the valence $4p$ electron results in the ground state of \ion{Sr}{4}, and ionization of the $4s$ sub-shell results in the configuration $4s^14p^6$ which is also bound in \ion{Sr}{4}. The computed cross sections are shown in Figure~\ref{fig:sr2_cs}. No measurements were available for comparison. Differences of order 30 - 40\% are observed between the Binary Encounter Dipole (shown) and Distorted Wave (not shown) single ionization cross sections. Given the accuracy of the Binary Encounter Dipole cross sections of \ion{Sr}{1} and \ion{Sr}{2} in the threshold regions, comparable accuracy for the BED approach in \ion{Sr}{3} is assumed.

The FAC calculations predict minor contributions from excitation autoionization from levels located just above the ionization potential. These levels take two forms: (1) doubly-excited states whose summed contribution is $<2\%$ at the peak of the cross section, and (2) contributions from singly excited states with configurations of the form $4p^5n\ell$, which take on larger but similarly negligible values. Estimates of the energies of these Rydberg-like states using quantum defects from Ref.~\cite{Persson_1972} places their energies close to but below the ionization potential. In total, EAI contributions in \ion{Sr}{3} are predicted to be negligible, and the total ionization cross section is computed as the sum of the two direct ionization channels.

As a system relatively free of indirect ionization channels, \ion{Sr}{3} presents a unique case to compare the configuration average and Lotz approaches. All three present methodologies - FAC level-resolved, FAC configuration average, and Lotz - have been shifted such that they include the known ionization potential for the ground states. Below the peak of the cross section, both FAC level-resolved and empirical Lotz (total) cross sections are similar but show sizable differences: the Lotz approximation overestimates both the $4s$ and $4p$ ionization cross sections. The CA cross section (here also without excitation autoionization) is larger than both the level-resolved and empirical cross sections, with differences decreasing in magnitude as $E$ approaches 5 keV. 

\subsection{Inner Shell Ionization}\label{subsec:innershells}
For \ion{Sr}{1}~-\ion{}{3} FAC calculations were pursued using the Distorted Wave approximation for inner shell ionization of all shells with binding energies below 5 keV. This corresponds to ionization of shells down to and including the $2s$ electron. The resulting cross sections are shown in Figure~\ref{fig:innershell_cs}. The configuration average cross sections are similar in both shape and magnitude to the level-resolved cross sections, which is interpreted as an insensitivity to many-body effects. Despite the large differences observed between FAC and Lotz single ionization cross sections, the Lotz approximation predicts inner shell cross sections close to the FAC calculations. For the dominant channels ($4s$ and $3d$), the cross section is modestly overestimated by the Lotz approximation.

At 5 keV, the sum of inner shell processes contributes 30-40\% of the total ionization cross section for \ion{Sr}{1}. For \ion{Sr}{2} and \ion{Sr}{3}, this is generally reduced to 20-30\%, depending on whether one considers FAC or Lotz cross sections. Among the inner shells only a handful contribute meaningfully to the total ionization cross section: for \ion{Sr}{1}, $4p$ ionization provides nearly 25\% of the total ionization cross section, and for \ion{Sr}{2} $3d$ ionization comes to dominate the multi-ionization contributions. Each shell's fractional contribution to the total ionization cross section is relatively constant as the ion stage increases. This implies that the single ionization cross section which comprises the bulk of the total cross section carries the most significant impact on the ionization properties of these ions, though inner shell ionization may lead to large changes in ion fractions due to multi-ionization effects. The impact of including inner shell ionization effects is explored later in this manuscript.

\section{Cross Section Dependence of Non-Thermal Ionization in Kilonovae}\label{sec:nonthermal}

\subsection{Spencer-Fano Implementation}
Beyond several days post-merger, the ejecta from a kilonova has sufficiently expanded that the role of non-thermal electrons - produced by beta decay within the radioactive ejecta - increases and can begin to compete with or dominate over photoionization~\citep{brethauer2025nonthermalionizationkilonovaejecta}. The energy of the non-thermal electrons is deposited into heating of the thermal electron population, and excitation and ionization of the ions in the ejecta. The distribution of non-thermal electron energies then naturally depends on the ionization and excitation properties of the ions. To probe the impact of updated ionization cross sections on the energy-dependent fluxes of non-thermal electrons solutions of the Spencer-Fano equation~\citep{Spencer1954,Kozma1992} were pursued. 

A methodology similar to that used for near-neutral Te ions in Ref.~\cite{Bromley_2026} is adopted, which is expanded here to explicitly consider both excitation autoionization and inner shell ionization. We use the Spencer-Fano modeling code \textsc{pynonthermal}~\citep{pynonthermal}, a Python package whose solver is similar to that of the ARTIS radiative transfer code~\citep{Shingles2019}. We use version 2026.9.30, which includes a self-consistent ionization balance solver, and handling of excitation autoionization and ionization channels that remove more than one electron.

\begin{figure*}
   \includegraphics[scale=0.6]{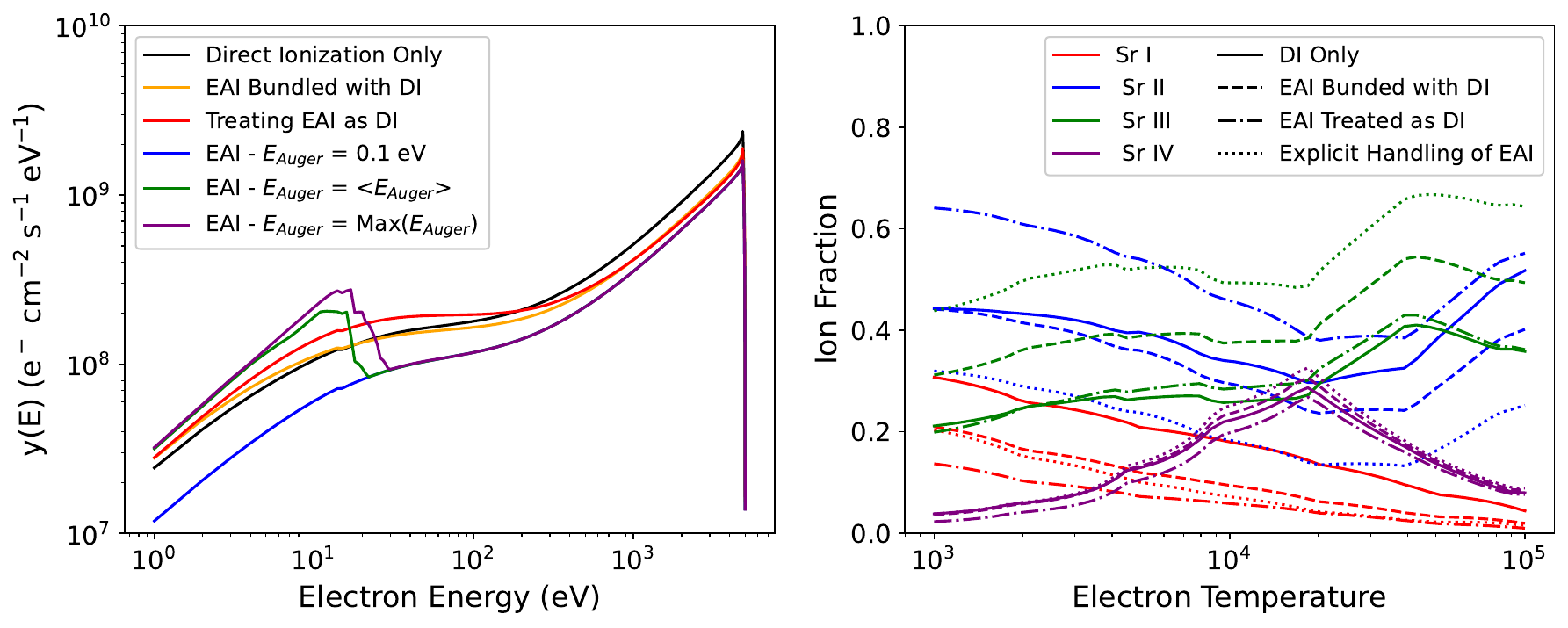}
\caption{(Left) Comparison of non-thermal electron degradation spectra at $T = 3100$~K, with all parameters equal except for different treatments of excitation autoionization: no EAI (black), using total cross sections produced by summing DI and EAI channels (orange), treating EAI as direct ionization (red), and explicit handling of excitation autoionization with all Auger electrons initialized at 0.1 eV (blue), the mean Auger energy for a given channel (green), and the maximum of the Auger energies for a given EAI channel (purple). (Right) Ionization balances as a function of thermal electron temperature using the same assumptions. Ionization balances are insensitive to the energies of the Auger electrons here, and only the curve utilizing the mean Auger energies is shown.}\label{fig:eai_comp}
\end{figure*}

Within the Spencer-Fano models electrons are injected uniformly over the top 1/30 of the energy grid, following Ref.~\cite{Kozma1992}. The maximum energy is here chosen as 5 keV to accommodate the inner shell ionization channels (down to and including $2s$), and so the source function spans 4.83 to 5 keV. The electrons are injected into a pre-defined plasma with an energy deposition rate of $q(t) = 0.15~(t/1.4~\textrm{days})^{-1.3}$ eV s$^{-1}$~ion$^{-1}$, where $t$ is the time since the merger measured in days, and 0.15 eV s$^{-1}$ ion$^{-1}$ represents a global average of the models of Shingles \textit{et al.}~\cite{Shingles2023} at 1.4 days post-merger. The deposition rate per unit volume is $q(t)\times{n}$, where $n$ is the number density of ions. A time of $t = 10$ days is chosen such that non-thermal ionization remains significant but the influence of photoionization by the radiation field is reduced. We assume a pure-strontium plasma, composed of \ion{Sr}{1} - \ion{Sr}{4}, undergoing spherical expansion. For ejecta with a total mass of $M = 0.05~M_\odot$ at velocity $v$ the mass density of the pure-Sr ejecta is estimated by $\rho = 3M/(4\pi(vt)^3)$. At $t$ = 10 days and a nominal velocity of $v = 0.1$c, the corresponding number density of Sr atoms is $n=9.37\times10^6$~cm$^{-3}$.

Within the Spencer-Fano model, the relative abundances of the four included charge states must also be determined. For a given electron temperature, recombination rates are computed from the rate coefficients of Sr ions reported in Ref.~\cite{McElroy2026}. The ion populations are then solved together with the Spencer-Fano equation. Let $\Gamma_{i,k}$ be the non-thermal rate coefficient (s$^{-1}$) for ionization of charge state $i$ by a channel that removes $k$ electrons. Recombination proceeds one charge state at a time with total rate coefficient $\alpha_{j+1}(T)$. In steady state, the net rate at which ions cross each boundary between adjacent charge states $j$ and $j+1$ is zero, which can be equated to
\begin{equation}\label{eq:ionbalance}
  \sum_{i \le j}\; \sum_{k > j-i} n_i\,\Gamma_{i,k} = n_{j+1}\, n_e\, \alpha_{j+1}(T).
\end{equation}
The above ionization balance includes only non-thermal ionization and the supplied recombination rates. Photoionization, thermal collisional ionization, and charge exchange are omitted. For single ionization only, this reduces to $n_i \Gamma_{i,i+1} = n_{i+1} n_e \alpha_{i+1}$ for each ion, and the free electron density follows from charge neutrality, $n_e = \sum_i q_i n_i$, where $q_i$ is the charge of state $i$. For given $\Gamma_{i,k}$, Eq.~\ref{eq:ionbalance} and charge neutrality determine the ion fractions and electron density. However, the rate coefficients depend on the degradation spectrum, which depends on the ion populations and on $n_e$ through the loss function. The solutions of both the fluxes of non-thermal electrons $y(E)$, and the ion fractions $n_i$, are therefore iterative. The Spencer-Fano equation is solved for trial populations, the $\Gamma_{i,k}$ are evaluated, and the ion fractions and $n_e$ are updated from Eq.~\ref{eq:ionbalance}. The rate coefficients are damped between iterations for stability. The iteration stops when every $\Gamma_{i,k}$ changes by a relative amount below $10^{-4}$. 


Within the models the energies of non-thermal electrons are tracked as they interact with the plasma, described by a thermal electron density $n_e$ and ions with populations $n_i$, through excitation, ionization, and heating. Each direct ionization event produces a spectrum of secondary electrons, whose energies are assumed to follow the distribution given in Opal \textit{et al.}~\cite{Opal1971} as 
\begin{equation}\label{eq:opal}
    P(E_p,E_s) = \frac{1}{J\textrm{arctan}[(E_p-I)/2J]}\frac{1}{[1+(E_s/J)^2]}
\end{equation}
where $E_p$ is the incident electron energy, $I$ is the ionization potential of the particular ionization pathway, and $J$ is a cutoff energy, here assumed to be $J=0.6I$. Electron impact excitation and ionization from these secondary electrons is included self-consistently.

\subsection{Excitation Autoionization}

When compiling total ionization cross sections it is common to sum the direct ionization and excitation autoionization cross sections. In the context of non-thermal electron degradation this approach is inappropriate, as EAI channels have higher thresholds and produce different distributions of secondary electrons. Excitation autoionization removes the excitation energy from the primary (incident) electron and produces Auger electrons at discrete energies. For the excitation autoionization channels of e.g. \ion{Sr}{1}, the Auger electrons span the energy range from $\sim$0 eV to $\sim$25 eV. The electrons at the higher end of this range can contribute to ionization, through both direct and indirect (EAI) ionization, and the lower energies solely to excitation/heating.

Within \textsc{pynonthermal}, an excitation autoionization channel is treated as follows. The primary electron loses exactly the excitation energy $E_{\rm exc}$ of the autoionizing level, as for a bound-bound excitation, and Eq.~\ref{eq:opal} is not used. One Auger electron is then injected at a single energy $E_{\rm A}$ via the source term in Eq.~8 of Ref.~\cite{Shingles2019}. The ion retains $E_{\rm exc} - E_{\rm A}$, and only this energy is counted as ionization. Energy conservation gives $E_{\rm A} = E_{\rm exc} - I$ for an ion left in its ground state, in which case the ion retains $I$. The channel contributes to the ionization rate coefficient and to the ionization balance as a single ionization.

For \ion{Sr}{1}, thousands of excitation and excitation autoionization channels contribute. To make their inclusion numerically approachable, we first compute the spectrum of Auger electrons, complete with fractional probabilities, for each excitation autoionization channel. Both excitation and excitation autoionization cross sections are then binned by their threshold energy onto the internal energy grid with spacing 1~eV. This binning of the electron impact excitation cross sections reduces the number of channels from $\sim$143000 to a more manageable value ($\sim$35). During this procedure the distribution of Auger electron energies is also weighted, summed, and binned on the internal energy grid, where comparable reductions in the number of excitation autoionization channels are also achieved. This binning procedure leads to at most 3\% changes in ion fractions between binned and unbinned cross sections for models at $T = 1000$~K, and decreasing with increasing temperature, and leads to a significant decrease in the computation time.

To probe the necessity of considering the \textit{spectrum} of Auger energies resulting from excitation autoionization, we first probe the impact of differing treatments of excitation autoionization on the degradation spectra, $y(E)$, and the resulting ionization balances. Figure~\ref{fig:eai_comp} (left) shows a comparison of degradation spectra computed for $T = 3100$~K, where the only differences are the treatment of excitation autoionization. Ignoring excitation autoionization completely (black curve) overestimates the fluxes of high energy electrons. Including EAI, by summing the corresponding cross sections with those of direct ionization, i.e. direct import of the total cross sections in Figs.~\ref{fig:sr0_cs} - \ref{fig:sr2_cs}, also leads to an overestimate of the high energy electron flux. If one separates out EAI and treats the process identically to direct ionization, complete with the Opal \textit{et al.} distribution of secondary electrons, there is again a decrease in fluxes at high energies and an increase at lower energies at and below the Auger injection energies ($E \lesssim 25$~eV).

\begin{figure}
\includegraphics[scale=0.8]{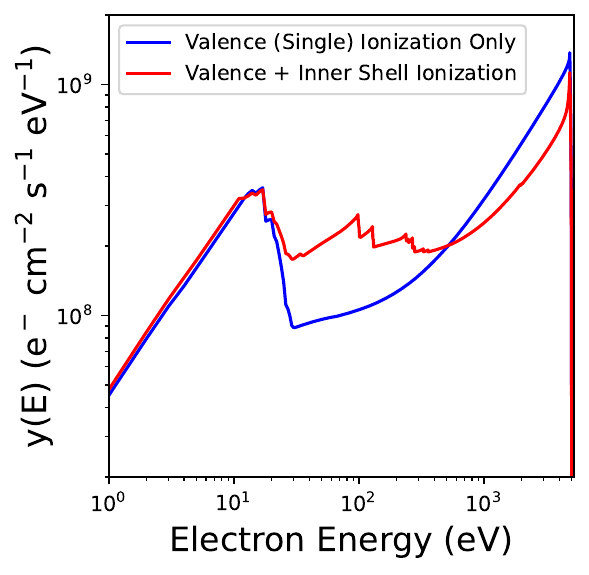}
\caption{Representative degradation spectra of non-thermal electrons in a pure-Sr plasma at 10 days post-merger for a thermal electron temperature of 1000~K. Both the free electron density and the ion fractions are solved self-consistently for each calculation. Single ionization, excitation autoionization, and electron impact excitation are handled identically between models. Calculations are shown excluding (blue) and including (red) inner shell ionizations up to and including removal of the $2s$ electron. All inner shell ionizations are assumed to lead to a double ionization.}\label{fig:edists}
\end{figure}

None of the approximate treatments of EAI as a direct ionization process matches the degradation spectrum found when handling excitation autoionization explicitly. Most importantly, the degradation spectra near and below the EAI threshold are only mildly affected by the energies of the Auger electrons. Above the EAI threshold(s), the fluxes of high-energy electrons are independent of the Auger energies. As a consequence of the insensitivity to the Auger electron energies in these cases, consideration of a \textit{spectrum} of Auger energies for each EAI channel is unnecessary, and minimal changes in ion fractions are introduced by assuming a single energy for the Auger electron of each EAI channel. 

Figure~\ref{fig:eai_comp} (right) shows the ionization balances resulting from these different treatments of excitation autoionization, as a function of electron temperature, where the ion fractions are solved self-consistently from the recombination rates and the ionization rates from the Spencer-Fano solver. None of the models which treat excitation autoionization as a direct ionization process produces ion fractions that agree with models explicitly treating excitation autoionization as excitation with the addition of an Auger electron. Notably, ion fractions of \ion{Sr}{2} and \ion{Sr}{3} - which are the end stages of ionization from \ion{Sr}{1} and \ion{Sr}{2}, both of which exhibit large contributions from EAI - differ by factors of $\sim$2 from models with explicit handling of EAI. In the remainder of this work, we treat excitation autoionization as an excitation process, with the addition of a single Auger electron injected at the mean energy of the Auger cascades for each threshold-binned EAI channel.

\subsection{Inner Shell Ionization}\label{innershellionization}
Inner-shell processes which result in a net double (or greater) ionization warrant comment. These multi-ionization events proceed as, at minimum, a two-step process. First an inner shell hole is formed and the distribution of secondary electrons depends on the ionization threshold of the inner shell~(Eq.~\ref{eq:opal}). Subsequent Auger cascades eject electrons at characteristic energies. Unless otherwise noted, we assume all inner shell ionizations lead to a net double ionization. The interactions between the Sr ions and these Auger electrons are included with the source term, found in Eq.~8 of Ref.~\cite{Shingles2019}. For a double ionization of charge state $i$ through a shell with threshold $I_{\rm shell}$, the ion retains the sum of the ground-state ionization potentials, $I_i + I_{i+1}$. The Auger electron is injected at the single energy $I_{\rm shell} - I_i - I_{i+1}$. This neglects fluorescence and excited final states, and is thus an upper limit on the Auger electron energy.

\begin{figure*}
        \centering
   \includegraphics[scale=0.65]{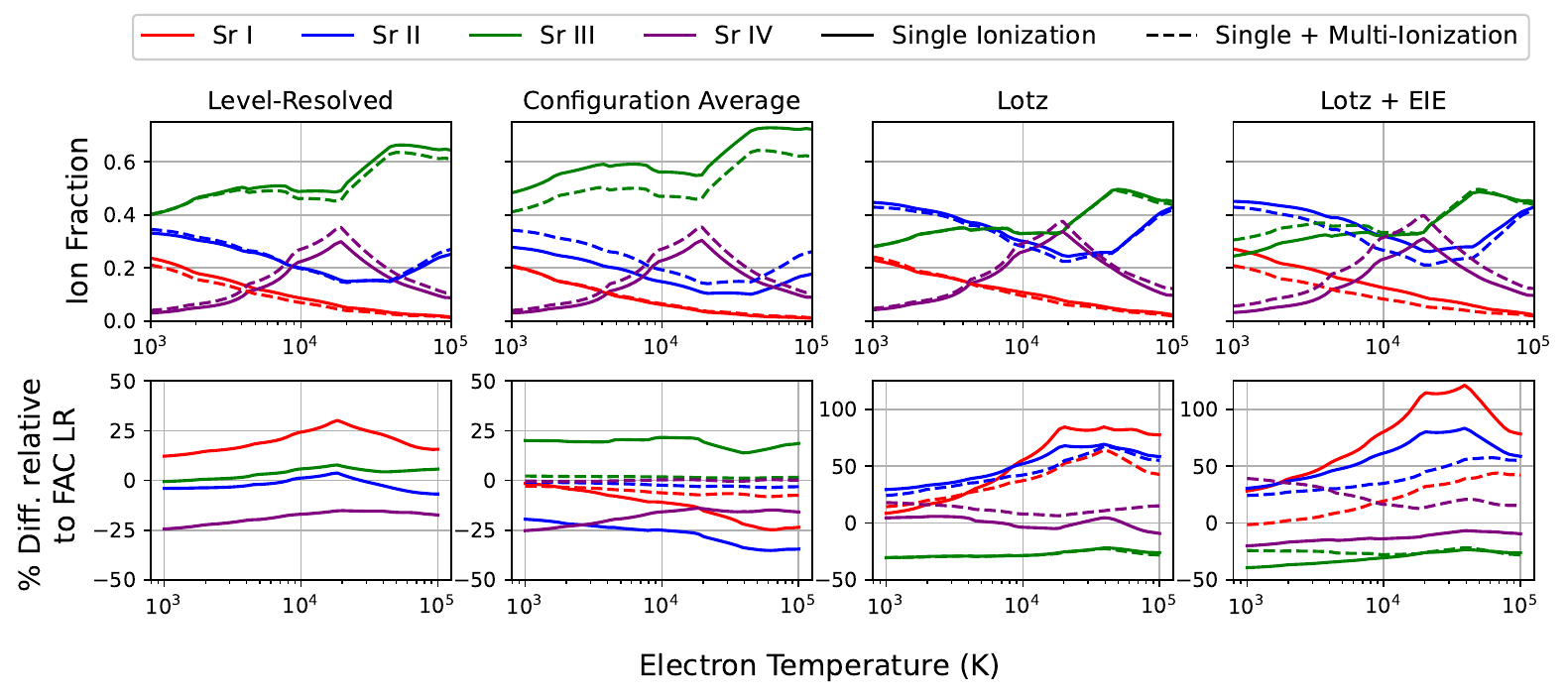}
\caption{(Top) Ion fractions computed from iterative solutions of the Spencer-Fano equations at each thermal electron temperature for different cross section datasets. All calculations utilized the same recombination data from Ref.~\cite{McElroy2026}. Ion fractions are shown for models without inner shell ionization (solid lines) and with inner shell ionization (dashed lines) for ions of strontium ($Z = 38$). Panels (left to right) show models using: FAC level-resolved (left), FAC configuration average (left middle), Lotz ionization only (right middle), and Lotz ionization with empirical electron impact excitation cross sections (right). (Bottom) Percent difference between each ionization balance model (`X') and the FAC level-resolved model with inner shell ionizations (`LR') computed as $100\times$(X-LR)/LR. Colors and linestyles in the bottom panel are identical to those in the top panel. Note that the bottom panels for the FAC-based models have different $y$ axis limits than the empirical models in order to distinguish differences in the ion fractions.}\label{fig:ibcomp_single_vs_multi}
\end{figure*}

Degradation spectra in representative models with and without inner shell ionizations are shown in Figure~\ref{fig:edists}. Despite comprising only 20 - 40\% of the total ionization cross sections (Fig.~\ref{fig:innershell_cs}), inner shell ionization has a dramatic impact on the fluxes of non-thermal electrons. Including inner shell ionization leads to a decrease in the flux of electrons above the inner shell ionization thresholds. Ionization through an inner shell channel removes a comparably larger energy from the non-thermal electron pool compared to an ionization that removes a valence electron. At the same time, the enhancement in the fluxes of non-thermal electrons with energies between 20 - 400~eV can be attributed to the increased energies of secondary electrons, which depend on the threshold of each ionization channel in the assumed Opal \textit{et al.} (Eq.~\ref{eq:opal}) distribution of secondary electrons.

\subsection{Full Model Results}
We now turn to quantifying the differences in ionization balances computed from different cross section datasets, including both excitation autoionization and inner shell ionization with the aforementioned treatments, in Spencer-Fano models of non-thermal electron degradation. For a fixed deposition rate at $t = 10$ days, we solved the Spencer-Fano equation and coupled ionization balance using four different cross section datasets: (1) the presently-computed FAC level-resolved excitation and ionization cross sections, (2) the presently-computed FAC configuration-average cross sections, (3) empirical models with only ionization cross sections predicted by the Lotz~\cite{Lotz1967} formalism, and (4) empirical models using Lotz ionization cross sections and approximated excitation cross sections using the formalism of van Regemorter~\cite{vanRegemorter1962}.

The resulting ionization balances are shown in Figure~\ref{fig:ibcomp_single_vs_multi}, which shows ion fractions (top row) and per cent differences from the FAC models using level-resolved cross sections for both valence and inner shell ionization (bottom row). Omitting inner shell ionization leads to up to $\pm$25\% changes in ion fractions compared to models with only single ionization. This trend is present across all four cross section datasets. In general, it is also found that the FAC configuration average cross sections produce self-consistent ion fractions within $\sim$5\% of those found in the FAC level-resolved model, despite the sizable differences in the valence ionization cross sections of both \ion{Sr}{1} and \ion{Sr}{2} (Figs.~\ref{fig:sr0_cs} - \ref{fig:sr1_cs}), particularly for \ion{Sr}{2} where the EAI cross sections in the CA approach are nearly a factor of three larger than their benchmarked level-resolved counterparts.

Both empirical models including and excluding electron impact excitation (`Lotz' and `Lotz + EIE') predict ion fractions within $\sim$25\% of the FAC-based models. Despite the increased direct ionization cross section partially offsetting the lack of excitation autoionization in the empirical cross sections, the ion fractions show larger deviations from the FAC level-resolved models when compared to the configuration average results. These differences are reduced when including inner shell ionization in the absence of excitation, but degrade once empirical excitation cross sections are included. Moreover, the empirical models reveal a trend wherein differences with the FAC-based models increase with thermal electron temperature.

\begin{figure*}
        \centering
   \includegraphics[scale=0.65]{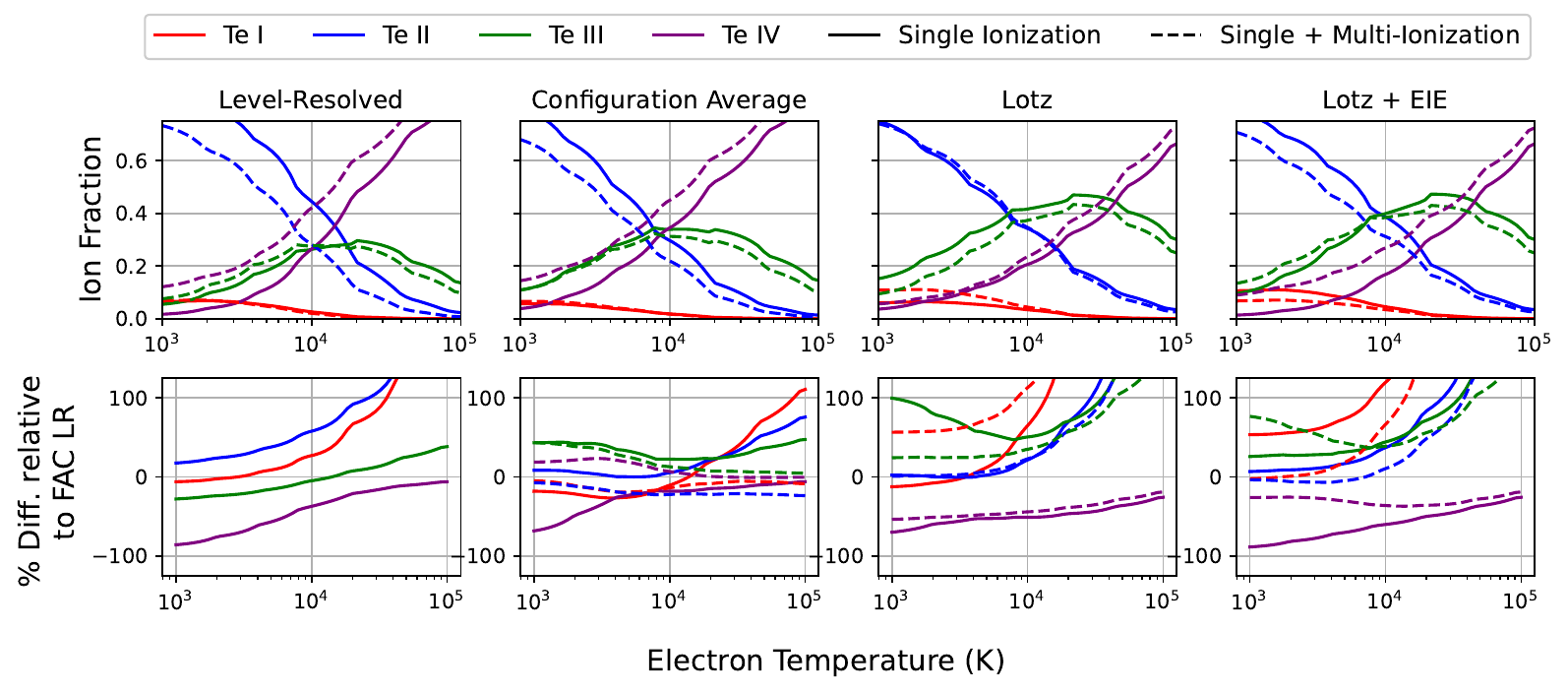}
\caption{(Top) Ion fractions computed from iterative solutions of the Spencer-Fano equations at each thermal electron temperature for different cross section datasets. All calculations utilized the same recombination data from Ref.~\cite{McElroy2026}. Ion fractions are shown for models without inner shell ionization (solid lines) and with inner shell ionization (dashed lines) for ions of tellurium ($Z = 52$). Panels (left to right) show models using the following cross sections: FAC level-resolved (left), FAC configuration average (left middle), Lotz ionization only (right middle), and Lotz ionization with empirical electron impact excitation cross sections (right). (Bottom) Percent difference between each ionization balance model (`X') and the FAC level-resolved model with inner shell ionizations (`LR') computed as $100\times$(X-LR)/LR. Colors and linestyles in the bottom panel are identical to those in the top panel.}\label{fig:ibcomp_single_vs_multi_te}
\end{figure*}

To investigate whether these comparisons are a fortuitous consequence of the cross sections for Sr ions, or a sign of a more general trend, we repeated the present analysis but for Te ions. We adopt the ionization and excitation cross sections reported in Ref.~\cite{Bromley_2026}, which we supplement with new calculations of the inner shell ionization cross sections. The same re-binning of both excitation and excitation autoionization cross sections, as well as binning of the Auger spectra, is also adopted.

The resulting ionization balance comparisons are shown in Figure~\ref{fig:ibcomp_single_vs_multi_te}. The shapes of the ion fractions, as a function of thermal electron temperature, are similar to those reported in the models of Ref.~\cite{Bromley_2026}. However, the present implementation includes self-consistent computation of the electron density and ion fractions, as well as explicit treatment of excitation autoionization, which is significant for \ion{Te}{3}. Similar trends found in Sr ions are again found for Te ions. First, omission of multi-ionization through inner shell ionization leads to $30\%$ or greater changes in ion fractions in the FAC level-resolved model. Configuration-average models with multi-ionization produce ion fractions of all four ion stages within $\sim$20\% of those predicted by the FAC level-resolved model. Both empirical models (with and without inner shell ionization and/or excitation) show increased scatter of the ion fraction differences with the FAC-based models. Above $\sim$5000~K these differences increase significantly, and depend on whether one includes empirical estimates of electron impact excitation.

\subsection{Extrapolation to Higher Ion Stages}
The large populations of both \ion{Sr}{4} and \ion{Te}{4} in the present models suggest that higher charge states should be included in the model. However, the ionization balance models presented in Figs.~\ref{fig:ibcomp_single_vs_multi} - \ref{fig:ibcomp_single_vs_multi_te} assume that inner shell ionization results in a double ionization event. This assumption simplifies the complexity of the inner shell ionization process, where subsequent Auger cascades can produce much higher ionization stages. Calculations by Kaastra and Mewe~\cite{Kaastra1993} probed the mean ion charges resulting from the production of inner shell holes in Fe ions; $K$-shell ionization of near-neutral Fe ions results in Auger cascades that produce a range of charge states with an average between $+4$ to $+6$. The formation of $L$ and $M$ shell holes in Fe ions produces charge states ranging from +3 to +6 and from +2 to +3, respectively. Similar calculations for neutral Pt ($Z = 78$) by Chaynikov \textit{et al.}~\cite{Chaynikov2025} reveal similar trends with larger average ionization stages. Though no inner shell Auger yields are available for neutral Sr, configuration-dependent Auger yields computed by Chaynikov \textit{et al.}~\cite{Chaynikov_Te} show that charge states exceeding Te$^{10+}$ can be produced from Auger cascades following the production of an inner shell hole in neutral Te. 

To probe the impact of modeling ionization processes more appropriately, a series of expanded Spencer-Fano/ionization balance calculations was performed for tellurium ions with a maximum charge state of +10. In the absence of ion-dependent recombination rates, we assume each of the added ion stages has identical recombination rate coefficients as \ion{Te}{3}. For ionization cross sections, we performed configuration average calculations in the Distorted Wave approximation. We assume for \ion{Te}{4}-\ion{Te}{9} that each inner shell ionization produces a net double ionization. 

The ionization balances resulting from this procedure are shown in Figure~\ref{fig:te_expanded} and compared to the CA models from Figure~\ref{fig:ibcomp_single_vs_multi_te}. Two sets of expanded models are shown. The first, shown as dotted lines, assumes all inner shell ionization processes result in a double ionization. This leads to negligible changes in the ion fractions of \ion{Te}{2}, but affects \ion{Te}{1} and \ion{Te}{3} at the 5-10\% level. A similar model (dashed lines) is also shown, where the inner shell ionization cross sections of \ion{Te}{1} - \ion{Te}{3} are converted to final-ion-resolved cross sections using the branching ratios reported in Ref.~\cite{Chaynikov_Te} for neutral Te. In this expanded model where ionizations greater than a double ionization are included, the energy $I_{shell} - I_i - I_{i+1}$ (see Sec.~\ref{innershellionization}) is assumed to be shared equally by the multiple Auger electrons. Given the similarities in the binding energies of the inner shells across this narrow range of ions, the branching ratios are not likely to change significantly as one increases the charge state from neutral Te to doubly ionized Te. However, this change leads to a significant (at or exceeding 20\%) reduction in the ion fractions of near-neutral Te ions. Similar changes in these and other ion fractions can be expected if the requisite branching ratios were available for higher charge states, and will also change as explicit ion-dependent recombination data becomes available.

Lastly, we note that the full model extrapolated to higher charge states shows that approximately 20 - 30\% of the Te atoms are found in charge states above triply ionized at the temperatures expected in KNe ejecta. Thus, spectral contributions from higher charge states may need to be considered, depending on the ionization properties of the element in question.

\section{Discussion}\label{sec:discussion}
Contributions from excitation autoionization are evident in the available measurements of the ionization cross sections of \ion{Sr}{1} and \ion{Sr}{2}. In the case of \ion{Sr}{1}, doubly excited states (below $\sim$21 eV) and excitations out of the closed $4p$ shell (above $\sim$21 eV) yield a relatively more complicated ionization cross section than direct ionization alone. Looking to \ion{Sr}{2}, the contribution of excitation autoionization out of the closed $4p$ shell is comparatively increased. Above the $4p$ excitation threshold in \ion{Sr}{2} over half of the total single ionization cross section results from indirect ionization channels. This inner shell `problem' is not unique to \ion{Sr}{2}. Badnell \textit{et al.}~\cite{Badnell1991} found similar complications in \ion{Ca}{2}, which sits just above Sr in the periodic table and has similar configurations (shifted down by one $n$ shell), and for which even close-coupling calculations faced challenges in finding agreement with measurement. Similar structures in the ionization cross sections of \ion{Rb}{1} were also identified by~Kim \textit{et al.}~\cite{Kim1998}. Though the cross section in the threshold region contributes negligibly to the non-thermal ionization rate, matching the measured cross section(s) in this region is critical for applications in kilonovae. For both electron impact ionization and electron impact excitation cross sections, which scale similarly at large energies as $\textrm{ln}(E)/E$, matching the magnitude of the indirect ionization channels near threshold directly alters the magnitude of the cross section at the higher energies carried by non-thermal electrons. Measurements of ionization cross sections near threshold and subsequent benchmarking of calculations against them is thus crucial for providing accurate ionization cross sections for use in kilonovae models. 

\begin{figure}[t]\includegraphics[scale=0.65]{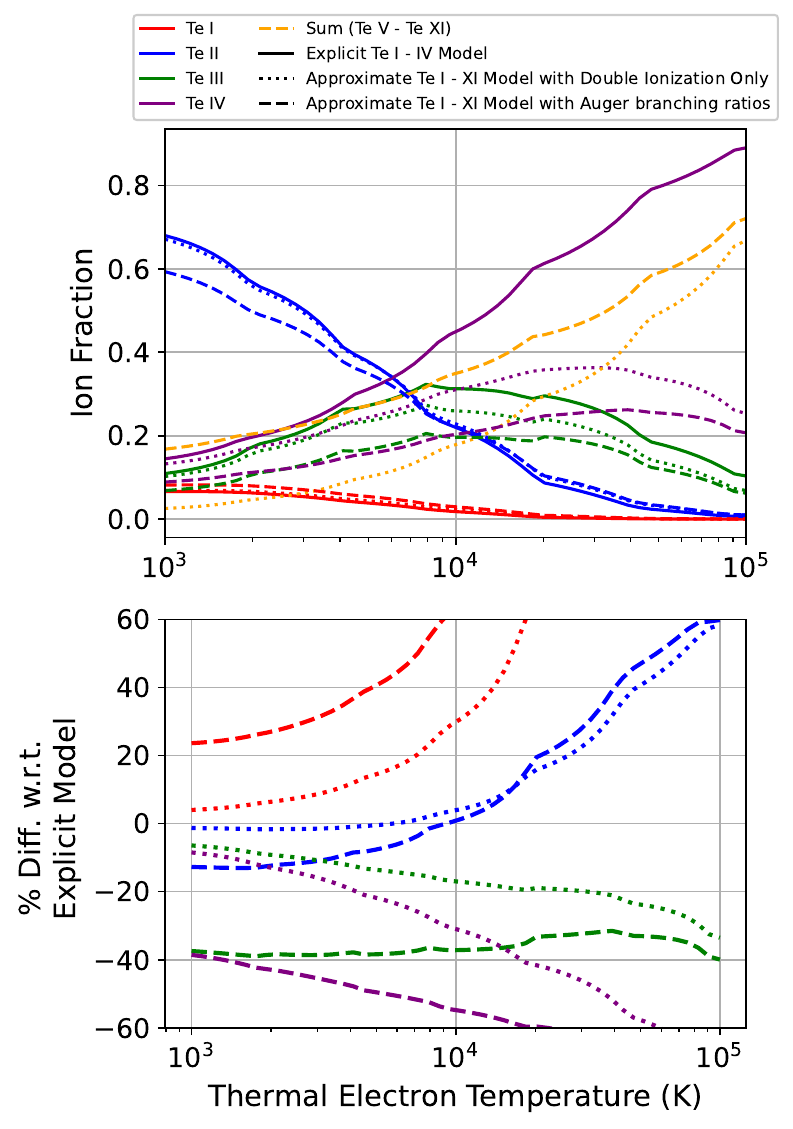}
\caption{(Top) Computed ionization balances from Spencer-Fano models using configuration average cross sections (identical to Figure~\ref{fig:ibcomp_single_vs_multi_te}) compared to  expanded ionization balances including up to \ion{Te}{11}. The expanded models use configuration average valence and inner shell ionization cross sections for all additional ions. Models treating inner shell ionizations as double ionization (dotted lines) are also compared to models where inner shell ionizations of \ion{Te}{1} - \ion{Te}{3} are converted to final-ion-resolved cross sections using the Auger branching ratios for neutral Te reported by Ref.~\cite{Chaynikov_Te}. Recombination data for \ion{Te}{5} and higher charge states are assumed to be identical to the data for \ion{Te}{4}. (Bottom) Per cent differences with respect to the explicit \ion{Te}{1} - \ion{Te}{4} model, computed as 100 * (Explicit - Expanded) / Explicit.}\label{fig:te_expanded}
\end{figure}

Comparing the present results to previous calculations on Te ions by Ref.~\cite{Bromley_2026}, the optimum user choices for producing agreement with measured cross sections is clearer for Sr ions. Optimization of the FAC's central field potential on the initial ion, as found for Te ions, also produces the greatest agreement with measurement for Sr ions. For Te ions the optimal user choice of either the Binary Encounter Dipole (BED) or Distorted Wave approximations in FAC calculations of direct ionization was ion-dependent. Though no measurements were available for comparison in the case of \ion{Sr}{3}, the agreement between calculation and measurement for \ion{Sr}{1} and \ion{Sr}{2} suggests that the BED approximation is possibly justified for \ion{Sr}{3}. 

While excitation autoionization was a minor contributor for Te ions - excepting the case of \ion{Te}{3} whose cross section exhibits structures analogous to those observed in \ion{Se}{3} by Ref.~\cite{Se2+_ionization} - here the contributions of excitation autoionization in \ion{Sr}{1} and \ion{Sr}{2} are readily apparent and unambiguous. The complexity of these structures provided stringent tests for the present calculations. The similarities between Sr ions and EAI structures identified in previous calculations for e.g. \ion{Ca}{2}~\cite{Badnell1991} suggest the same approaches used for Sr ions may be applicable to other elements and ions with similar structures. Considering commonly noted ions of interest to kilonovae, the present procedures might also be useful in studies of Ba ions as well as \ion{Y}{3}, \ion{La}{3}, and \ion{Cs}{1}.

Within the present Spencer-Fano and ionization balance models at kilonova-like plasma conditions, it was found to be necessary to treat the excitation autoionization cross sections of \ion{Sr}{1} and \ion{Sr}{2} as being akin to electron impact excitation, with the addition of an Auger electron. However, the resulting ionization balances are insensitive to the energies of the Auger electrons in the present cases. Despite large differences in the excitation autoionization cross sections computed at level-resolved and configuration-average resolution with the FAC, the resulting ionization balances agree at the 5-10\% level. In contrast, models built on empirical cross sections show sensitivity to the inclusion or exclusion of inner shell ionization and electron impact excitation. Once electron impact excitation is included in the empirical models, ion fractions show larger differences with the present FAC-based models. We note that for the purposes of non-thermal electron degradation, energy-dependent \textit{cross sections} are the required atomic inputs. In systems with complex level structures such as lanthanides, or systems with a large number of contributing excitation autoionization channels (such as \ion{Sr}{2}), distribution, storage, and inclusion of the large number of energy-dependent cross sections appears tenebrous. The present calculations adopted a procedure where both the excitation and excitation autoionization cross sections were binned by the threshold energy of each channel. This process lead to a significant reduction in the number of stored cross sections from over 10$^5$ channels to an approachable $\sim$35, and resulted in changes in computed ion fractions at the few per cent level at most. 

Comparison across ions, i.e. comparing the ionization balances of Sr ions and Te ions reported in Figs.~\ref{fig:ibcomp_single_vs_multi} - \ref{fig:ibcomp_single_vs_multi_te}, the agreement between FAC level-resolved and FAC configuration-average models agree at the 5 - 20\% level. These models included inner shell ionization, which was approximated as yielding a net double ionization. Expanding the model to higher ion stages, assuming all inner shell ionizations lead to only a double ionization, initially leads to negligible changes in ion fractions. However, explicit calculations of Auger cascades and their initial-configuration-dependent ion yields by Ref.~\cite{Chaynikov_Te} show that ionization of inner shells can produce net triple, quadruple, or higher ionization events. Considering the timescales over which these Auger cascades occur ($t\sim 10^{-10}$ seconds,~\cite{Chaynikov_Te}), the Auger branching ratios can be used to convert inner shell ionization cross sections into final-ion-resolved ionization cross sections for each shell. An expanded grid of ionization balance models built on these ratios (Fig.~\ref{fig:te_expanded}) shows further changes in ion fractions of near-neutral Te ions of order 20 - 50\%. Despite the present benchmarking of the excitation autoionization and inner shell ionization effects on ionization balances of Sr and Te ions, large uncertainties on ion fractions remain due to these unexplored Auger cascades. Future calculations of these ion- and configuration-dependent branching ratios would be valuable for refining ionization balance models used to study kilonovae ejecta.

\section{Conclusions}\label{sec:conclusions}
New calculations of the valence and inner shell ionization cross sections of \ion{Sr}{1}, \ion{Sr}{2}, and \ion{Sr}{3} were reported. For both \ion{Sr}{1} and \ion{Sr}{2} significant contributions from excitation autoionization were apparent in the computed single ionization cross sections and agree well with measurement~\cite{Vainshtein1972,Okudaira1970,Peart1975,Peart1989}. As electron impact excitation decreases as $\textrm{ln}(E)/E$ at high energies, near-threshold excitation autoionization directly contributes also at the higher electron energies that are primarily responsible for ionization by non-thermal electrons in kilonovae. Measurements of ionization cross sections near threshold are shown to be particularly useful for the benchmarking of theoretical calculations. In \ion{Sr}{3} contributions from excitation autoionization were found to be negligible, though there exists a sizable uncertainty of order 30\% in the single ionization cross section stemming from whether the Binary Encounter Dipole or Distorted Wave approximation is most appropriate. 

The cross section calculations were extended to inner shell ionization within the perturbative Distorted Wave formalism. Though single (valence) ionization consistently dominates the total ionization cross section, multi-ionization can contribute between 20 - 40\% of the total cross section. In all cases, the configuration average (CA) approximation was found to produce cross sections comparable to the level-resolved data. 

The computed cross sections were incorporated into a coupled Spencer-Fano and ionization balance model. Owing to the large contributions from excitation autoionization in \ion{Sr}{1} and \ion{Sr}{2}, it is shown that excitation autoionization should not be treated equivalently to direct ionization, nor should it be omitted. A grid of calculations shows that omission of excitation autoionization in \ion{Sr}{1} and \ion{Sr}{2}, wherein the process composes $\sim$half of the total ionization cross sections, leads to at times a factor of two changes in computed ion fractions. Rather, the excitation-like nature of the process, with the addition of Auger electrons, should be adopted and can have large impacts on degradation spectra and ion fractions. The Auger electrons ejected after excitation autoionization and inner shell ionization are included in the degradation spectra, each at a single energy. Negligible differences in ion fractions are found when considering either the mean or maximum Auger energy of each cascade channel. This treatment may need refinement for plasmas with high-$Z$ elements such as Pt ($Z = 78$), where Auger electrons may carry significantly higher energies.

Non-thermal electrons in kilonovae carry sufficiently large energies that inner shell holes can be formed. Omission of inner ionization  within the Spencer-Fano models results in changes in ion fractions exceeding at-times 50\%. Once included - with the approximation that each inner shell ionization yields a net double ionization - the FAC-based models agree well and show large differences from models based on empirical cross sections. The sizable ion fractions of \ion{Sr}{4} and \ion{Te}{4} suggest that higher ion stages should be included. An expanded model, extended up to \ion{Te}{11} with explicitly-computed ionization cross sections but approximate recombination data, shows minor changes in ion fractions when assuming inner shell ionizations lead to only a double ionization. However, conversion of near-neutral inner shell ionization cross sections into final-ion-resolved cross sections leads to changes in near-neutral ion fractions of order 20-50\%. These large changes can be attributed to uncertain or absent branching ratios for the final ion yields of successive Auger cascades following inner shell hole production, and the lack of recombination data for higher charge states. These unknown branching ratios are identified as a potentially significant source of uncertainty in the present ionization balance models, and should be constrained for other elements that are spectroscopically relevant to nebular kilonovae.

Future calculations should continue to investigate the ionization properties of near-neutral, high-$Z$ elements of interest to $r$-process nucleosynthesis. In particular, investigations into the ionization cross sections should consider the roles of excitation autoionization and Auger cascades from inner shell holes. Such efforts would be useful for developing models to interpret future observations of kilonovae, particularly in the nebular phase for which the James Webb Space Telescope is well suited to capture infrared spectral signatures of $r$-process elements.

The level-resolved direct ionization, excitation autoionization, and excitation cross sections  (the latter two binned over threshold energies) for \ion{Sr}{1}~-\ion{}{3} will be provided on Zenodo \href{https://doi.org/10.5281/zenodo.23779187}{here}. with supporting documentation. The data are stored in plain-text machine-readable tables with accompanying documentation describing the units and file contents.

\section*{Acknowledgments}
The authors gratefully acknowledge funding support from the National Science Foundation (Awards 2308012 and 2308013). The authors express their gratitude to the High Performing Computing team at Auburn University for their computational resources, including access to the Easley computing cluster.
{LJS and CPB are funded by the European Union (ERC, HEAVYMETAL, 101071865). Views and opinions expressed are, however, those of the authors only and do not necessarily reflect those of the European Union or the European Research Council. Neither the European Union nor the granting authority can be held responsible for them.}

\appendix

\bibliographystyle{elsarticle-num}
\bibliography{pt}






\end{document}